\documentclass[prl,aps,floats,twocolumn,superscriptaddress,floatfix]{revtex4}

\usepackage{amsmath}
\usepackage{graphics, graphicx}
\usepackage{epsfig}
\usepackage{ulem}

\def\be{\begin{equation}}
\def\ee{\end{equation}}
\def\beq{\begin{eqnarray}}
\def\eeq{\end{eqnarray}}
\def\qf{{}}

\begin{document}

\title{The No-Boundary Measure  of the Universe}

\author{James B. Hartle}
\affiliation{Department of Physics, University of California, Santa Barbara,  93106, USA}
\author{S.W. Hawking}
\affiliation{DAMTP, CMS, Wilberforce Road, CB3 0WA Cambridge, UK}
\author{Thomas Hertog}
\affiliation{Laboratoire APC, Universit\'e Paris 7, 10 rue A.Domon et L.Duquet, 75205 Paris, France {\em and}\\
International Solvay Institutes, Boulevard du Triomphe, ULB -- C.P. 231, 1050 Brussels, Belgium}

\bibliographystyle{unsrt}

\begin{abstract}

We consider the no-boundary proposal for homogeneous isotropic closed universes with a cosmological constant and a scalar field with a quadratic potential. In the semi-classical limit, it predicts classical behavior at late times if the initial scalar field is more than a certain minimum. If the classical late time histories are extended back, they may be singular or bounce at a finite radius. The 
no-boundary proposal provides a probability measure on the classical solutions which selects inflationary histories but is heavily biased towards small amounts of inflation. This would not be compatible with observations. However we argue that the probability for a homogeneous universe should be multiplied by $\exp(3N)$ where $N$ is the number of e-foldings of slow roll inflation to obtain the probability for what we observe in our past light cone. This volume weighting is similar to that in eternal inflation.  In a landscape potential, it would predict that the universe would have a large amount of inflation and that it would start in an approximately de Sitter state near a saddle-point of the potential. The universe would then have always been in the semi-classical regime.

\end{abstract}


\maketitle

\noindent{\bf Introduction}\\
The string theory landscape is believed to contain a vast ensemble of stable and metastable vacua that includes some with a small positive effective cosmological constant and the low energy effective field theory of the 
Standard Model. But the landscape by itself does not explain why we are in one vacuum rather than in some other.  For that one has to turn to cosmology and to a theory of the quantum state of the universe. 

A manifest feature of our quantum universe is the wide range of epoch and scale on which the laws of classical physics apply, {\it  including classical spacetime.} Classical spacetime is a prerequisite for the construction of effective theories, for cosmology, and for eternal inflation. But classical spacetime is not a property of every state in quantum gravity. Rather it emerges only for certain quantum states.

We calculate the probability measure on classical spacetimes predicted by the no-boundary wave function (NBWF) \cite{Hartle83}  to leading semiclassical order for homogeneous and isotropic minisuperspace models with a cosmological constant and a scalar field with a quadratic potential. We find the NBWF severely restricts the possible classical universes and argue that such classicality restrictions would act as a strong vacuum selection principle in the string landscape.

The NBWF predicts the probabilities of entire classical histories. But we are interested in the probability for our observations which are restricted to a (thickened) light cone located somewhere in the universe and extending over roughly a Hubble volume \cite{Hawking06}. To calculate such probabilities we must sum the probabilities for classical histories over all those that contain our data at least once \cite{Page97,Hawking07}. This defines the probability for our data in a way that is gauge invariant and dependent only on information in our past light cone. We will argue that the resulting probabilties favor an inflationary past and, in a landscape potential, suggest a semiclassical origin.

\noindent{\bf Classical Prediction in Quantum Cosmology} \\
In quantum cosmology states are represented by wave functions on the superspace of three-geometries and spatial matter field configurations. For the homogeneous, isotropic  models considered here minisuperspace is spanned by the scale factor $b$ and the value $\chi$ of the homogeneous scalar field. Thus, $\Psi = \Psi(b,\chi)$. 

The no-boundary wave function \cite{Hartle83} is defined by the sum-over-histories
\begin{equation}  
\Psi(b,\chi) =  \int_{\cal C} \delta g \delta \phi \exp(-I[a(\tau),\phi(\tau)]/\hbar) .
\label{nbwf}
\end{equation}
Here, $a(\tau)$ and $\phi(\tau)$ are the histories of the scale factor and matter field and $I[a(\tau),\phi(\tau)]$ is their Euclidean action. The sum is over cosmological geometries that are regular on a manifold with only one boundary at which $a(\tau)$ and $\phi(\tau)$ take the values $b$ and $\chi$. The integration is carried out along a suitable complex contour ${\cal C}$ which ensures the convergence of \eqref{nbwf} and the reality of the result. 

For some regions of minisuperspace the integral in \eqref{nbwf} can be approximated by the method of steepest descents. Then the wave function will be well approximated  to leading order in $\hbar$ by a sum of terms of the form 
\begin{equation}
\Psi(b,\chi) \approx  \exp\{[-I_R(b,\chi) +i S(b,\chi)]/\hbar\} ,
\label{semiclass}
\end{equation}
one term for each extremizing history.  The functions $I_R(b,\chi)$ and $-S(b,\chi)$ are the real and the imaginary parts of the action evaluated at the extremum. 
 In simple cases these extremizing histories may describe the nucleation of a Lorentzian spacetime by  a Euclidean instanton. But in general they  will be complex --- ``fuzzy instantons''.  
 
In order for wave functions of the form \eqref{semiclass} to predict an ensemble of Lorentzian histories with high probabilities  for classical correlations in time  further conditions must be satisfied.  A necessary one is  the {\it classicality constraint}
\begin{equation} 
|(\nabla S)^2| \gg |(\nabla I_R)^2|,
\label{classconstraint}
\end{equation}
where gradients and inner products are defined with the minisuperspace metric. 
When \eqref{classconstraint} holds the action $S$ satisfies the Lorentzian Hamilton-Jacobi equation. The NBWF then predicts the corresponding ensemble of Lorentzian histories. Their probabilities are $\exp[-2 I_R(b,\chi)]/\hbar]$ to leading order in $\hbar$. 

Two key points should be noted:
(1) The no-boundary wave function provides probabilities for entire classical {\it histories}. (2) The histories in the classical ensemble are not the same as the extremizing histories that provide the steepest descents approximation to the integral \eqref{nbwf}. The classical histories are real and Lorentzian and may have two large large regions. The extrema are generally complex  with only one large region.
\\
\noindent{\bf Scalar Field Model}\\
We have applied this prescription for classical prediction to homogeneous isotropic closed universes with a cosmological constant $\Lambda$ and a scalar field $\Phi$ with a quadratic potential $V(\Phi) = (1/2)m^2 \Phi^2$. 
We write the complex homogeneous isotropic metrics that provide the steepest-descent approximation to the no-boundary path integral \eqref{nbwf} as
\begin{equation}
ds^2=(3/\Lambda)\left[d\tau^2 +a^2(\tau) d\Omega^2_3\right].
\label{emetric}
\ee
The Euclidean action $I$ then takes the form
\beq
I[a(\tau),\phi(\tau)] & =& \frac{9\pi}{4\Lambda}\int_{{C}(0,\upsilon)}  d\tau \left[ -a {\dot a}^2 -a +a^3 
\right.\cr
& & \left. \qquad \qquad +a^3 \left({\dot\phi}^2 + \mu^2\phi^2\right)\right] .
\label{eact}
\eeq
We use units where $\hbar=c=G=1$ and define the measures $\phi=(4\pi/3)^{1/2} \Phi$ and $\mu \equiv  (3/\Lambda)^{1/2}m$. 
The contour  ${C}(0,\upsilon)$ in the complex $\tau$-plane connects the South Pole $\tau=0$ with an endpoint $\tau=\upsilon$ where $a$ and $\phi$ take real values $b$ and $\chi$.

We  evaluated the NBWF in the semiclassical approximation \eqref{semiclass}  by numerically solving the Friedman-Lema\^itre equations for each value of $b$ and $\chi$ along a suitable complex contour ${ C}(0,\upsilon)$. This gives complex analytic functions $(a(\tau), \phi(\tau))$ which are an extemum of the action. 
The value of the action at an extremum gives $I_R(b,\chi)$ and  $S(b,\chi)$.

The integral curves of $S$ are the classical solutions when the classicality constraint \eqref{classconstraint} is satisfied. The relation between position and momenta that follows from $S$ means that, in the semiclassical approximation,  the NBWF  predicts non-zero probabilities only for a one-parameter ensemble of the two-parameter family of  classical histories. Classical histories not in the ensemble have zero probability. The relative probabilities for histories in this classical ensemble are given by  $\exp[-2I_R(b,\chi)/\hbar]$  in the leading semiclassical approximation. These are constant along the integral curves. It is convenient to take $\phi_0 \equiv |\phi(0)|$ to be the parameter labeling different histories in this classical ensemble. 

\begin{figure}[t!]
\includegraphics[width=3in]{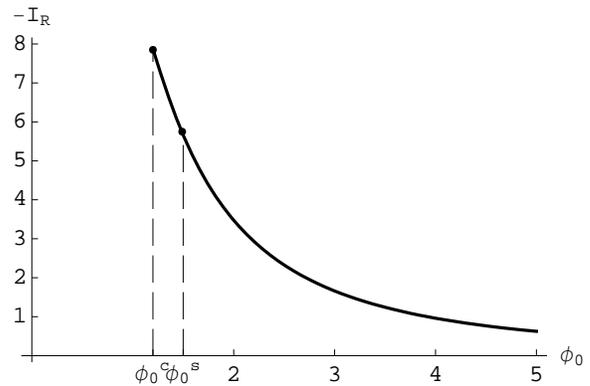} 
\caption{The values of $I_R$  of the one-parameter set of classical histories predicted by the no-boundary proposal in  a quadratic potential minisuperspace model with $\mu=3$ and $\Lambda =.03$. There are no classical histories for $\phi_0$ below a critical value $\phi_0^c$ at about $1.2$. The universe therefore requires a minimum amount of matter to behave classically at late times. A critical value $\phi_0^s$ at about $1.5$ separates large $\phi_0$ histories that bounce at a finite radius when extrapolated back from singular histories for smaller $\phi_0$.}  
\label{action}
\end{figure}

The classicality constraint \eqref{classconstraint} is not satisfied for all integral curves of $S$.  {\it Specifically, in the interesting regime where  $\mu>3/2$ we find the NBWF requires the universe to contain a minimum amount of scalar field energy at early times to behave classically at late times.} (The value of $\mu$ based on today's $\Lambda$ would be very much larger.) From now on we restrict to this range.  Similar conclusions were reached in \cite{GR90} for the $\Lambda=0$ case.
This is illustrated in Figure \ref{action}, where we show $I_R(b,\chi)$ for all members of the ensemble of classical histories predicted by the NBWF in a $\mu=3$ model with $\Lambda \approx .03$. There is a critical value $\phi_0^c$ below which there are no classical solutions. The lower bound $\phi_0^c$ implies a lower bound on the scalar field in the corresponding classical histories. The critical value $\phi_0^c$ increases slightly with $\mu$ and tends to 1.27 when $\Lambda \rightarrow 0$,  for fixed  $m$. 

The classicality constraint is closely related to the slow roll condition of scalar field inflation. We make this precise in Figure \ref{inflation} where we plot the trajectories in $(H,\phi)$ variables where $H$ is the instanteous Hubble constant $H=\dot b/b$. We show  five members of the ensemble of classical histories in the $\mu=3$ model for $\phi_0$ between 1.3 and 2. When we follow the histories back in time to higher values of $H$, they all lie within a very narrow band around $H =\mu\phi$. But this is precisely the regime that corresponds to slow roll inflationary solutions, as emphasized recently in  \cite{Gibbons06}. Hence the NBWF plus classicality at late times {\it implies} inflation at early times. 
 
\begin{figure}[t]
\includegraphics[width=3in]{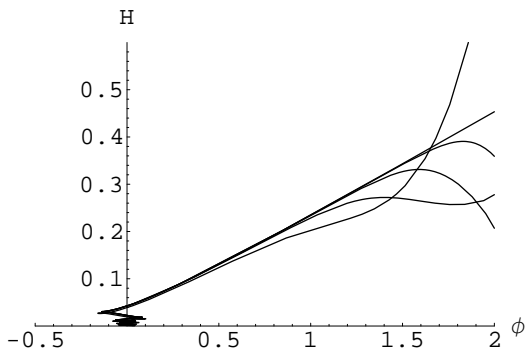} 
\caption{The no-boundary wave function predicts that all histories that behave classically at late times undergo a period of inflation at early times as shown here by the linear growth of the instantaneous Hubble constant $H$ in five representative $\mu=3$ classical histories.}
\label{inflation}
\end{figure}

For $\phi_0$ smaller than a critical value $\phi_0^s > \phi_0^c$ the allowed classical histories of the universe are singular in the sense that their matter densities exceed the Planck density. But for $\phi_0 > \phi_0^s$ they bounce at a finite radius in the past. This is possible despite the singularity theorems because a scalar field and the cosmological constant  violate the strong energy condition. Near the bounce the universe approaches a de Sitter state with radius $\sim (\mu \phi_0)^{-1}$. Such non-singular solutions form only a small subset of all scalar field gravity solutions but have  significant probability in the no-boundary state.

Even for the histories in the ensemble that are classically singular at an early time the NBWF unambiguously predicts probabilities for late time observables such as CMB fluctuations, because it predicts probabilities for histories rather than their initial data. The existence of singularities in the extrapolation of some classical approximation in quantum mechanics is not an obstacle to prediction but merely a limitation of the validity of the approximation. 
Indeed, there could be quantum mechanical transitions rather than classical ones across cosmological singularities that connect two classical regimes \cite{Turok07}.

Individual classical bouncing histories are not generally time-symmetric about the bounce, although the time asymmetry is small for large $\phi_0$. However, the reality of the NBWF implies the ensemble of allowed classical histories is time symmetric. For every history in this ensemble, its time reverse is also a member.

For the universes that bounce at a minimum radius it seems likely that the NBWF will predict that fluctuations away from homogeneity and isotropy will be at a minimum at the bounce  and grow away from the bounce for at least a while on either side (cf. \cite{HLL93}). This means that the thermodynamic arrow of time is likely to point away from the bounce on either side of it.  Events on one side are therefore unlikely to have a causal impact on the other and have much explanatory value.  
This is very different from the causality in pre-big bang universes where the arrow of time points in one direction throughout the spacetime. \\
\noindent {\bf Top Down Cosmology}\\
The NBWF gives the probabilities of entire classical histories. But we are interested in probabilities that refer to our data which are limited to a part of our past light cone. Among these are the top-down probabilities \cite{Hawking06} for our past conditioned on our present data. These are obtained by summing over the probabilities for classical spacetimes that contain our data at least once, and over the possible locations of our light cone in them. 
 
These sums can be implemented concretely in our closed,  homogeneous, isotropic minisuperspace models as follows: Approximate the probability for our data on the past light cone by the probability of data in  a Hubble volume on an appropriate surface of homogeneity. Assume that our data are otherwise detailed enough that they occur only once on this surface \cite{HS07}. The sum is then over the spatial locations of our Hubble volume in that surface of homogeneity in all classical spacetimes that last {\qf sufficiently long}.

The classicality constraint $\phi_0>\phi_0^c$ implies that all histories in the classical ensemble inflate (Figure 2). {\qf The condition that the universe lasts $\sim$14 Gyr further restricts the ensemble, requiring 
$\phi_0$ to be larger than a critical value $\phi_0^g > \phi_0^c$.} On average each history has the same behavior shortly after inflation ends and thus predicts the same observable physics for every Hubble volume at the present time. But the classical histories differ in the value $\phi_i \approx \phi_0$ of {\qf the inflaton at the start of inflation}, and consequently in the volume of the present surface of homogeneity. None of these properties is directly observable and should be summed over. The sum over our location therefore multiplies the NBWF probability for each classical spacetime in the ensemble by the number of Hubble volumes in the total present volume --- a factor proportional to $\exp(3N)$. This favors larger universes and more inflation. In a larger universe there are more places for our Hubble volume to be. 

Volume weighting increases the probability of a large number of efoldings. For quadratic potentials with realistic values of  $m$ and $\Lambda$ the constraints of classicality and minumum age yield a restricted ensemble of histories whose volume weighted probabilities {\qf slightly} favor a large number of efoldings \cite{Hartle07} that are necessary for explaining the observed spatial flatness.
An important feature of the volume weighted probability distribution is that there is a wide region where the probability is strongly increasing with $N$. The gradient of the probability distribution $ \sim \exp(3N-2I_R)$ with respect to  $\phi_i$  is positive if
\be \label{eternal}
V^3 \geq \vert V_{,\phi} \vert ^2
\ee
which, intriguingly, is the same as the condition for eternal inflation \cite{Vilenkin83,Page97}.

 Hence, there is a striking contrast between the unconditioned  bottom-up probabilities that favor small amounts of inflation and the top-down probabilities conditioned on our data that  favor larger amounts.

\noindent{\bf Landscape Potentials}\\
\begin{figure}[t]
\includegraphics[width=1.65in]{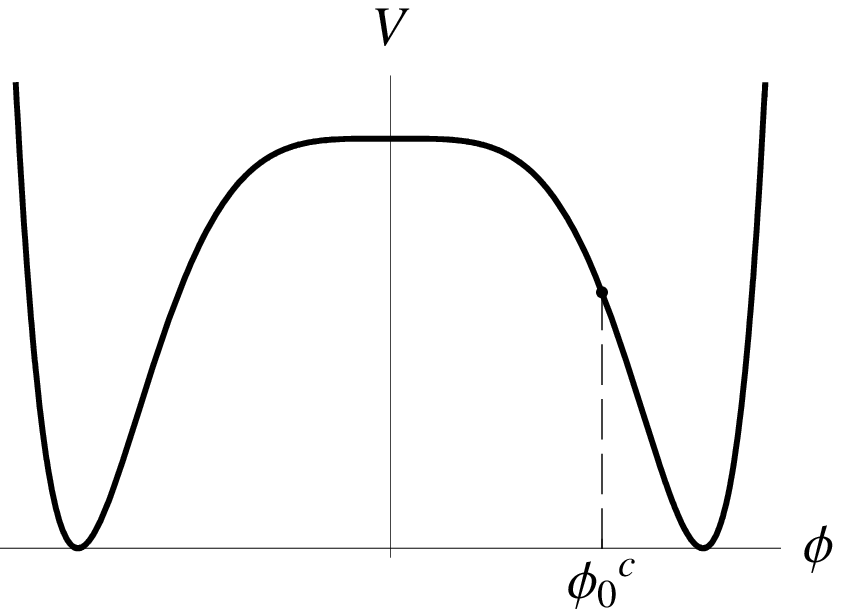} \hfill
\includegraphics[width=1.65in]{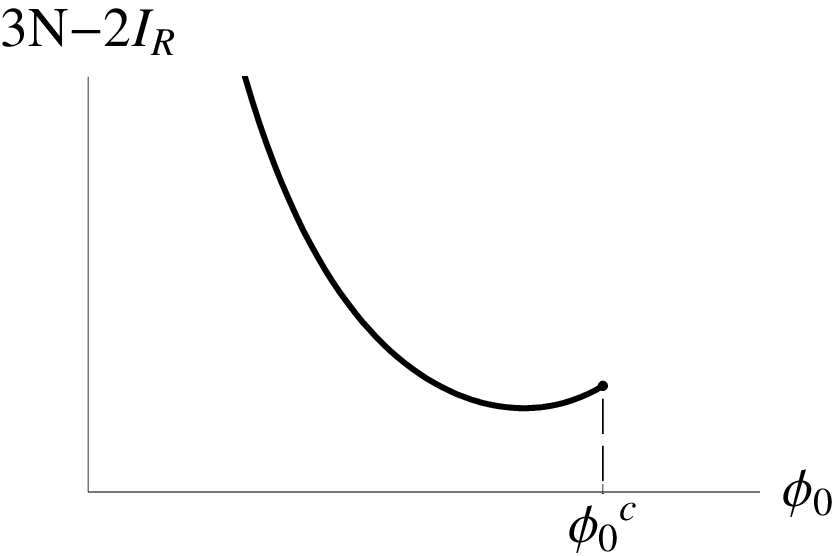}
\caption{To account for the different possible locations in the universe of the Hubble volume that contains  our data one ought to multiply the no-boundary amplitudes by a volume factor.
In regions of the landscape around a maximum of the potential (left), we expect this to have a significant effect on the probability distribution over $\phi_0$ and hence over $N$ (right). The effect of  a classicality constraint is also shown.}
\label{volume}
\end{figure}
A typical landscape potential will have several saddle-points besides the quadratic directions discussed above. For saddle points with more than one descent direction, there will generally be a lower saddle-point with only one descent direction, and with lower action.  If this descent direction is sharply curved we expect the classicality constraint \eqref{classconstraint} not to be satisfied in analogy with  the  case of quadratic potentials. Hence the no-boundary amplitude for universes that emerge from around such saddle-points will be approximately zero. Thus only broad saddle-points with a single descent direction will give rise to significant no-boundary amplitudes for universes that behave classically at late times. Only a few of the saddle-points will satisfy the demanding condition that they be broad, because it requires that the scalar field varies by order the Planck value across them. {\it The classicality constraint, therefore, acts as a vacuum selection principle.} 

By analogy with quadratic potentials we expect the classical histories predicted by the NBWF for $\phi_0$ near a broad maximum of $V$ to have an early period of inflation, during which the scalar field rolls down to a nearby minimum of $V$. (We assume for simplicity that all vacua are consistent with the Standard Model.) As before, the no-boundary proposal favors a small number of efoldings, i.e. histories where $\phi_0 \approx \phi^c_0$ (see Fig \ref{volume}a).
However in contrast with quadratic potentials, near a broad maximum the volume factor more than compensates for the reduction in amplitude due to the higher value of the potential. The resulting probabilities of past histories consistent with present data significantly favor a large number of efoldings. This is illustrated in Figure \ref{volume}b  and  discussed in \cite{Hartle07}. 

This leads us to predict that in a landscape potential, the most probable homogeneous history of the  universe that is consistent with our data started in an unstable de Sitter like state near a broad saddle-point of $V$. Because the dominant saddle-points are well below the Planck density we expect the most probable histories are bouncing solutions of the field equations which lie entirely in the semi-classical regime. They have a large amount of slow roll inflation. During this the scalar field evolves from the saddle-point to the neighbouring minima of $V$, populating only a few of the possible vacua in the landscape. 
 \\ \noindent{\bf Inhomogeneities}\\
In this paper we have discussed homogeneous universes only. However, one can also consider inhomogeneous perturbations. It appears that the volume weighting can overcome the gradient action for very long wavelength perturbations that leave the horizon while \eqref{eternal} is satisfied. This suggests the NBWF with volume weighting will predict a universe that is very inhomogeneous on very large scales. Eternal inflation \cite{Vilenkin83} also  predicts large scale inhomogeneities but the connection, if any, with this picture is not yet clear to us. In any event no additional `measure' would be needed  to derive  the probabilities for this structure. The NBWF in principle provides that.

\noindent{\bf Acknowledgments} We thank Neil Turok for stimulating discussions. The work of JH was supported in part by the National Science Foundation under grant PHY05-55669.

\nopagebreak

\end{document}